\documentclass[twocolumn,twoside,slac_two]{revtex4}
\usepackage{graphicx}
\usepackage{fancyhdr}
\begin{document}

\title{Study of the X-ray emission mechanism of radio-loud narrow-line Seyfert 1 galaxy}


\author{H. Shirakawa}
\affiliation{School of Science, Hiroshima University, 1-3-1, Kagamiyama, Higashi-hiroshima, Hiroshima, Japan 739-8526}
\author{Y. Fukazawa, Y. Tanaka, R. Itoh and K. Kawaguchi}
\affiliation{School of Science, Hiroshima University, 1-3-1, Kagamiyama, Higashi-hiroshima, Hiroshima, Japan 739-8526}

% In case co-authors are many, please use the following expression.

%\author{F. Author$^1$, S. Author$^2$}

%\affiliation{$^1$University/Institute, City, State, Postal Code Country}
%\affiliation{$^2$Colalborative University/Institute, City, State, Postal Code Country}

\begin{abstract}
1H0323+342 is one of narrow-line radio-loud Seyfert 1 galaxies (RL-NLS1), which is a new class of gamma-ray emitting AGNs. Narrow-line Seyfert 1 galaxies (NLS1) have a small-mass black hole, but its mass accretion rate is almost as high as Eddington limit. Therefore, by observing NLS1s, we can know the evolution of supermassive black holes at the center of galaxies. Some of NLS1s are radio-loud and we call them RL-NLS1. From past observations, multi-wavelength spectrum of RL-NLS1s is similar to that of typical blazars; the synchrotron emission in the lower energy band up to the optical band, and inverse Compton scattering of low energy photons from disk, torus, and broad line region. X-ray band is a transittion region between the synchrotron and inverse Compton, and also there is a possible disk/corona emission.
Therefore, we studied the energy-dependence of time variability of the X-ray emission of 1H0323+342, which have been observed by Suzaku in 2009 and 2013, in order to constrain the emission mechanism. We found that the lower energy below 1 keV and the higher energy above 7 keV show a different variability from the middle energy band, indicating at least two emission components in the X-ray band. X-ray spectrum is not a simple power-law, but requires an additional features; a broken power-law plus flat hard component, or a power-law plus a relativistic reflection component. Each spectral component seems to vary independently.
\end{abstract}

\maketitle

\thispagestyle{fancy}


\section{Nallow-line radio-loud Seyfert 1 galaxy (RL-NLS1)}

Narrow-line Seyfert 1 galaxy (NLS1) is a class of active galactic nuclei
(AGN). The width of optical emission lines is narrower than that of
Seyfert 1 galaxies, and NLS1s do not exhihit strong X-ray absorption
like Seyfert 2 galaxy. In the X-ray spectrum, there is often a large soft-excess. NLS1s are identified by the following three characteristics.
\begin{enumerate}
 \item FWHM (H$\beta$) $<$ 2000 kms$^{-1}$ (Goodrich 1989)
 \item $[$OIII$]$ / H$\beta <$  3 (Osterbrock \& Pogge 1985)
 \item strong permitted Fe II emission lines (Boroson \& Green 1992)
\end{enumerate}
NLS1s have a small-mass black hole, but its mass accretion rate is almost 
as high as Eddington limit (Marconi et al. 2008). We can study the 
evolution of supermassive black holes at the center of galaxies by 
observing NLS1s. 
\subsection{Nallow-line radio-loud Seyfert 1 galaxy (RL-NLS1)}
Most of NLS1s are radio-quiet (R $<$ 10, R : radio loudness, ratio 
of 5 GHz radio to B-band flux densities) (Kellermann et al. 1989), 
but ~7 $\%$ of NLS1s are radio-loud (R $>$ 10) and ~2.5 $\%$ are 
very radio-loud (R $>$ 100) (Komossa et al. 2006), and they are 
called as radio-loud narrow-line Seyfert 1 galaxies (RL-NLS1).

With early Fermi observation, 
GeV gamma-ray emission has been discovered from PMN J0948+0022, 
one of RL-NLS1s (Abdo et al. 2009a).
After that, GeV gamma-ray emission was found from other three RL-NLS1s (Abdo et
al. 2009b), so it is found that RL-NLS1s generally emit GeV gamma-rays.
From past observations, the multi-wavelength spectrum of RL-NLS1 is
similar to that of typical blazars; the synchrotron emission in 
the lower energy band up to the optical band and the inverse Compton 
scatterred X-ray and gamma-ray emission of low energy photons from disk, torus, and broad line 
region. X-ray band is a transition region between the synchrotron 
and the inverse Compton, and also there is a possible disk/corona emission.

As described above, X-ray emission mechanism of RL-NLS1s is
uncertain. Therefore, we studied the energy-dependence of time
variability of the X-ray emission of a RL-NLS1s 1H 0323+342, which have been observed by Suzaku in 2009 and 2013.

\section{Data analysis}
\subsection{Energy-dependence of time variability}

\begin{figure}[htbp]
\centering
\includegraphics[width=1.0 \linewidth]{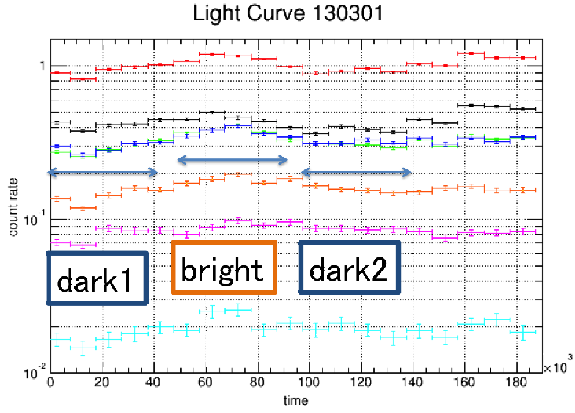}
\caption{Suzaku X-ray light curves of 1H 0323+342 in varios energy bands. Band1 :
 0.5--1 keV (black), band2 : 1--2 keV (red), band3 : 2--3 keV (green),
 band4 : 3--5 keV (blue), band5 : 5--7 keV (orange), band6 : 7--10 keV
 (magenta), band7 : Fe-k line region (cyan). Definition of bright and
 dark period is also shown for spectral analysis.} \label{lc13}
\end{figure}

\begin{figure}[htbp]
\centering
\includegraphics[width=1.0 \linewidth]{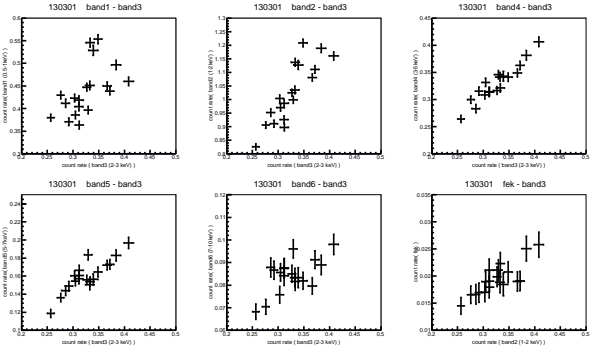}
\caption{Correlations of count rates between band3 (2--3 keV) and other energy bands in 2013} \label{multi}
\end{figure}

Fig.\ref{lc13} is a light curve of 1H 0323+342 observed by Suzaku in
2013. 1H 0323+342 varies with a time scale of
several ks, but the soft X-ray band below 1 keV shows a independent
behavior from the other bands . Fig.\ref{multi} shows correlations of
count rates between 2--3 keV and other bands. The lower energy below 2 keV and the higher energy above 7 keV show a different variability from the middle energy band, suggesting that there are at least two spectral components in the X-ray band.

\subsection{Spectral fitting}

\begin{figure}[htbp]
\centering
\includegraphics[width=1.0 \linewidth]{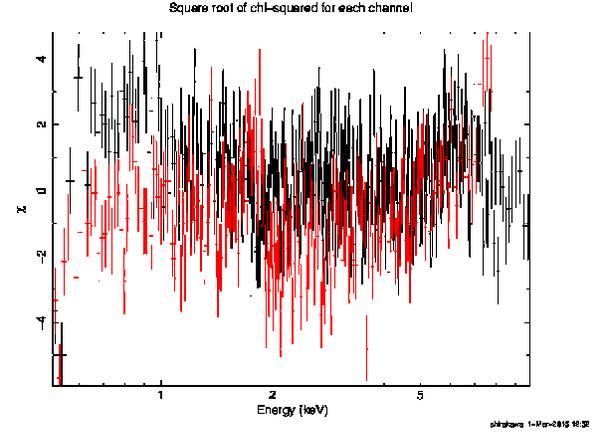}
\caption{Redisuals of fitting of the 1H 0323+342 spectra in 2013 with
 the absorved power-law model.} \label{redisual}
\end{figure}

First, we fit the X-ray spectrum of 1H 0323+342 with the absorbed
power-law model, but there remain redisuals at the higher and lower
energy region ($\chi^2 / dof$ = 1017.26/528) and thus the spectrum is not a simple power-law shape (Fig. \ref{redisual}).

\subsubsection{Jet emission model}
\begin{figure}[htbp]
\centering
\includegraphics[width=1.0 \linewidth]{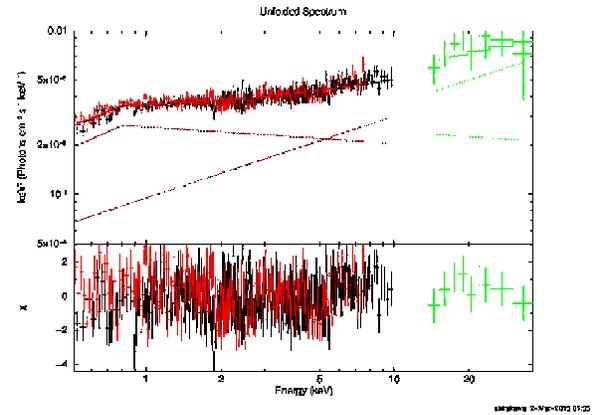}
\caption{Fitting the 2009 spectrum with a broken power-law plus a
 power-law model (jet emission model).} \label{bkn09all}
\end{figure}

\begin{figure}[htbp]
\centering
\includegraphics[width=1.0 \linewidth]{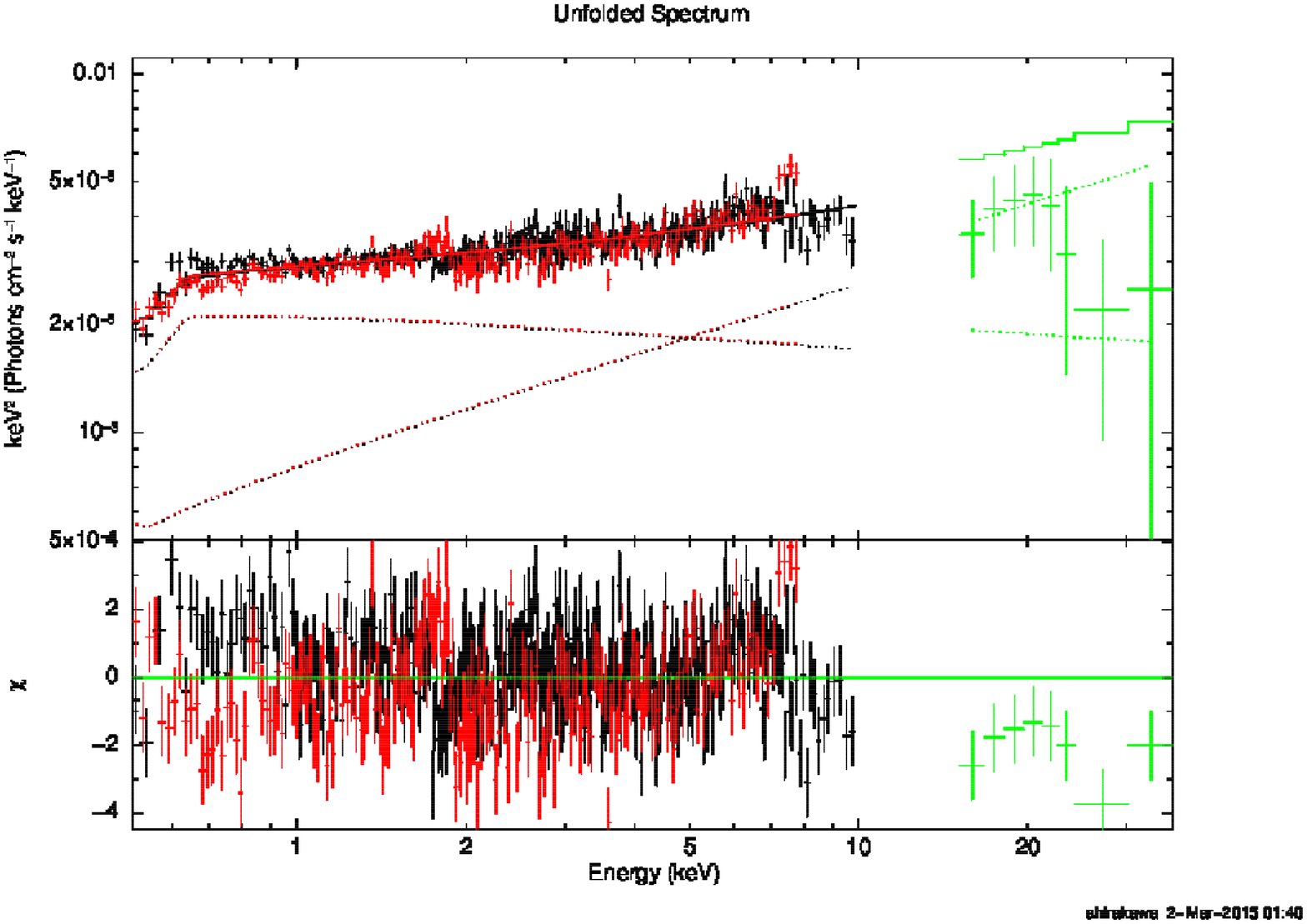}
\caption{Fitting the 2013 spectrum  with a broken power-law plus a
 power-law model (jet emission model).} \label{bkn13all}
\end{figure}

To reduce the redisual in Fig.\ref{redisual}, we added a broken
power-law model This model of one hard power-law and one broken
power-law represents a jet emission of inverse Compton and synchrotron, 
respectively. 
As in Fig.\ref{bkn09all} and Fig.\ref{bkn13all}, X-ray spectra can be fitted with this model for 2009 and 2013 (Table I). The breaking energy becomes around 0.7 keV.

\begin{table}[htbp]
\begin{tabular}{|c|c|c|c|c|c|}\hline
 &photon&breaking&photon& photon& $\chi ^2 /dof$ \\ 
&index 1& energy (keV)&index 2&index& \\ \hline
2009 &1.3$\pm$0.1&0.80$\pm$0.03&2.1(fix)&1.5(fix)&602.10/496 \\ \hline
2013 &0.33$\pm$0.36&0.63$\pm$0.02&2.1(fix)&1.5(fix)&975.22/536\\ \hline
\end{tabular}
\caption{Fitting parameters of jet emission model in 2009 and 2013} \label{paraeter1}
\end{table}

Next, we analyzed the spectra of three periods during the 2013
observation as in Fig.\ref{lc13}, defined by brightness in 2013. 
Only a broken power-law component varied, suggesting a fast variable
synchrotron emission if the jet emission model is correct.
This behavior is similar to blazars.

\subsubsection{Disk emission model}

\begin{figure}[htbp]
\centering
\includegraphics[width=1.0 \linewidth]{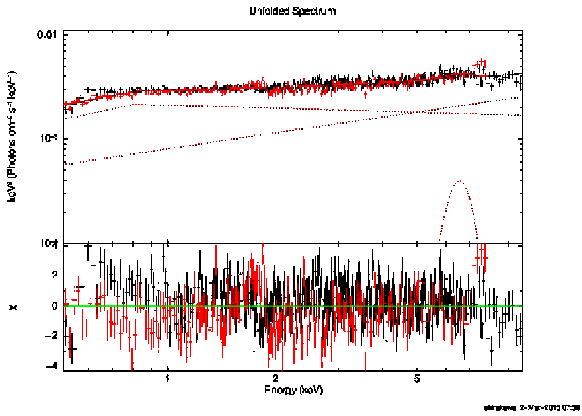}
\caption{Fitting with an additional Fe-K line (E=6.5keV, width=0.5keV), together with the above jet emission model.} \label{13poline}
\end{figure}

When we look at the residual fitted with jet emission model in 2013
(Fig.\ref{bkn13all}) in detail, there is a feature like a broad Fe line around 6 keV.
Therefore, we fit the 2013 spectrum with an additional Fe-K line (E = 6.5 keV, width = 0.5 keV), together with the above jet emission model (Fig.\ref{13poline}).
Fe line intensity is (1.2$\pm$0.4)$\times10^{-5}$ counts/s/cm$^2$
(2.7$\sigma$ statistical significance) and $\chi ^2 /dof = $ 908.28/528;
the fit improved.

\begin{figure}[htbp]
\centering
\includegraphics[width=1.0 \linewidth]{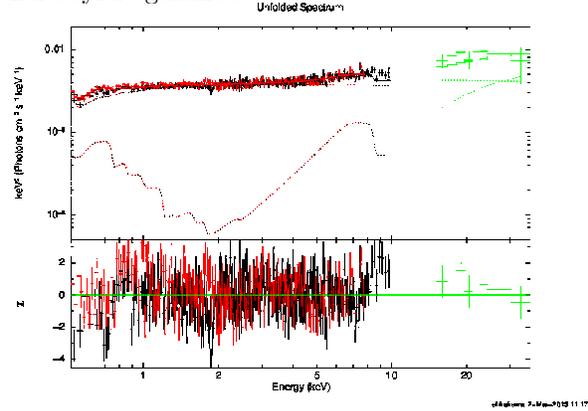}
\caption{Fitting the 2009 spectrum with disk emission model. See the
 text in detail.} \label{09kerall}
\end{figure}

\begin{figure}[htbp]
\centering
\includegraphics[width=1.0 \linewidth]{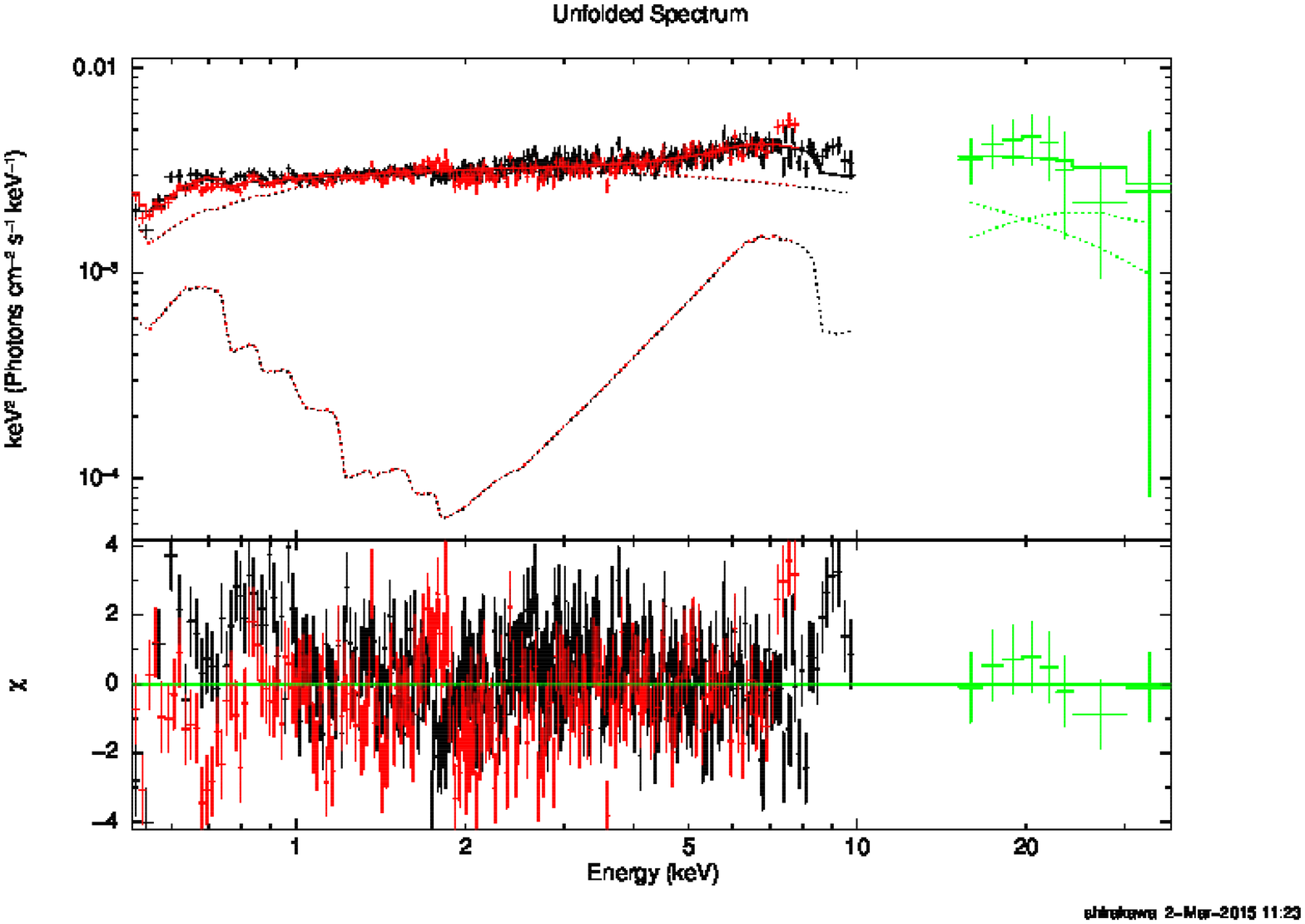}
\caption{Fitting the 2013 spectrum with disk emission model. See the
 text in detail.} \label{13kerall}
\end{figure}

Then, we try to fit the spectra with a single power-law model plus a
relativistic reflection model.
The relativistic refelction component represents a reflection of a
power-law incident photons by the ionized accretion disk around a rotating black hole.
We call this disk emission model.
In this case, the spectra are also be fitted (Fig.\ref{09kerall} and
Fig.\ref{13kerall}). 
The broad Fe line feature is represented by the relativistic reflection component.
The broken-like feature around 0.7 keV is 
represented by the Fe-L line complex. 

\begin{table*}[htbp]
\begin{tabular}{|c|c|c|c|c|c|c|c|c|}\hline
&cut-off energy(keV)&fold energy (keV)&photon index&a&Inclination (degree)&Fe/solar&Xi&$\chi ^2 /dof$ \\\hline
2009&0.1(fix)&1000(fix)&2.02$\pm$0.02&0.998$\pm$0.02&84.3$\pm$1.4&4.55$\pm$1.00&10.0(fix)&640.69/493\\\hline
2013&0.1(fix)&20(fix)&1.92$\pm$0.017&0.998$\pm$0.014&84.6$\pm$1.07&5.00$\pm$1.25&10.0(fix)&984.40/534\\\hline
\end{tabular}
\caption{Fitting parameters of disk emission model in 2009 and 2013} \label{parameter2}
\end{table*}

We also try to fit the spectra of two periods defined in fig \ref{lc13}
with disk emission model, as jet emission model. 
As a result, a "powerlaw" component varied. 
If disk model is correct, the variability is almost attributed to the 
disk/corona component and the reflection component is stable. 
This behavior is similar to other Seyfert galaxies.

\section{Sammary and Future works}

We suggests that X-ray emission of 1H 0323+342 has at least two emission
components, based on energy-dependence of time variability. If the X-ray
emission of 1H 0323+342 is dominated by jet emission, a variable
component is a synchrotron, and it is similar to other blazers. If the emission is dominated by disk/corona emission, disk/corona emission varies while disk reflection is stable, this is similar to Seyfert galaxies.
We cannot distinguish these two models by current data. Therefore an
extensive study by ASTRO-H observations with high energy resolution and
good sensitivity in wide X-ray band is hopeful.


\end{document}